\documentclass{PoS}

\def\slash#1{\!\not\!\!#1}

\title{Eigenvalue distribution of the Dirac operator at finite temperature with (2+1)-flavor dynamical quarks using the HISQ action}

\ShortTitle{Eigenvalue distribution of the Dirac operator at finite temperature with (2+1)-flavor $\dots$} 

\author{\speaker{H.~Ohno},$^a$\thanks{Current address: Fakult\"{a}t f\"{u}r Physik, Universit\"{a}t Bielefeld, D-33501 Bielefeld, Germany}\hspace{0.5em}
	U.M.~Heller,$^b$ F.~Karsch,$^{c,d}$ and S.~Mukherjee$^d$ \\
        \llap{$^a$}Graduate School of Pure and Applied Sciences, University of Tsukuba, \\
        Tsukuba, Ibaraki 305-8571, Japan \\
        \llap{$^b$}American Physical Society, \\
        One Research Road, Ridge, NY 11961, USA \\
        \llap{$^c$}Fakult\"{a}t f\"{u}r Physik, Universit\"{a}t Bielefeld, \\
	D-33501 Bielefeld, Germany \\
        \llap{$^d$}Physics Department, Brookhaven National Laboratory, \\
	Upton, NY 11973, USA \\
        E-mail: \email{ohno@het.ph.tsukuba.ac.jp}}

\abstract{We report on the behavior of the eigenvalue distribution of the Dirac operator in (2+1)-flavor QCD at finite temperature, using the HISQ action. 
	We calculate the eigenvalue density at several values of the temperature close to the pseudocritical temperature.
	For this study we use gauge field configurations generated on lattices of size $32^3 \times 8$ with two light quark masses 
        corresponding to pion masses of about 160 and 115 MeV.
        We find that the eigenvalue density below $T_c$ receives large contributions from near-zero modes which become smaller as 
	the temperature increases or the light quark mass decreases. Moreover we find no clear evidence for a gap in the eigenvalue density up to 1.1$T_c$.
        We also analyze the eigenvalue density near $T_c$ where it appears to show a power-law behavior consistent with what is expected in the
	critical region near the second order chiral symmetry restoring phase transition in the massless limit.}

\FullConference{XXIX International Symposium on Lattice Field Theory \\
                 July 10 - 16 2011\\
                 Squaw Valley, Lake Tahoe, California}

\begin{document}

\section{Introduction\label{intro}}

The chiral phase transition is one of the most important features of quantum chromodynamics (QCD) in our understanding of the properties
of strongly interacting matter in a hot medium. QCD with $N_f$-flavors of massless quarks has an $SU_L(N_f)\times SU_R(N_f)$ chiral symmetry,
which is spontaneously broken in the vacuum, {\it i.e.,} at zero temperature.
The vanishing of the chiral condensate $\langle \bar{\psi}\psi \rangle$,
an order parameter of the chiral phase transition, signals the restoration of chiral symmetry at high temperatures.

It was shown by Banks and Casher \cite{Banks-Casher} that the eigenvalue density of the Dirac operator $\rho(\lambda)$ is related to $\langle \bar{\psi}\psi \rangle$
by $|\langle \bar{\psi}\psi \rangle| = \pi \rho(0)$ in the limits of infinite volume and vanishing quark mass.
Therefore low-lying eigenvalues of the Dirac operator play a key role in the breaking and restoration of chiral symmetry.
In particular, $\rho(0)$ is nonzero in the chirally broken phase and $\rho(0)$ vanishes when the chiral symmetry is restored.
Whether $\rho(\lambda)$ develops a gap around $\lambda=0$ above the critical temperature $T_c$ is another important subject
since it is related to restoration of the $U_A(1)$ symmetry. The $U_A(1)$ symmetry is explicitly broken by the anomaly which exists independently of temperature.
Thus the $U_A(1)$ symmetry is expected to remain broken above $T_c$. If the difference of isovector susceptibilities $\omega\equiv \chi_P-\chi_S$ \cite{Chandrasekharan:1998yx},
where $\chi_i\equiv \int d^4 x \langle j^{k}_i(x)j^k_i(0)\rangle$ $(i=P,S)$, $j^k_P(x)\equiv \bar{\psi}(x)\tau^k i\gamma_5\psi(x)$ and $j^k_S(x)\equiv \bar{\psi}(x)\tau^k \psi(x)$,
is considered, a relation
\footnote{Here and in the following we ignore any explicit dependence of $\rho(\lambda)$ on the light sea quark mass $m$, i.e. we write $\rho(\lambda)\equiv \rho(\lambda,m)$.
This will eventually become important also for a discussion of the critical behavior.},
\begin{equation}
\omega=\int^{\infty}_0 d\lambda \frac{4m^2\rho(\lambda)}{(\lambda^2+m^2)^2} \ ,
\end{equation}
where $m$ is a quark mass, can be obtained in a manner similar to the Banks-Casher relation.
Therefore, if $\rho(\lambda)$ has a gap around $\lambda=0$, then $\omega=0$ in the chiral limit, which means $\chi_P=\chi_S$ and that the $U_A(1)$ symmetry is effectively restored.

If the chiral phase transition occurs continuously and $\rho(\lambda)$ changes smoothly as the temperature increases,
a power-law behavior of $\rho(\lambda)$ such as $\rho(\lambda) \sim \lambda^{\alpha}$ becomes important because $\alpha$ is related to critical
exponents of the phase transition. It has been suggested that for $N_f=2$, there is a second order phase transition \cite{Pisarski-Wilczek} belonging
to the same universality class as 3-dimensional $O(4)$ spin models. In fact, for staggered quarks, although only an $O(2)$ symmetry is preserved rather than
$O(4)$, at least $O(N)$ scaling has been suggested in an analysis of the magnetic equation of state \cite{Ejiri:2009ac,Kaczmarek:2011zz}.
Similarly, studies performed with Wilson quarks indicated $O(4)$ scaling \cite{Iwasaki:1996ya,Ali Khan:2000iz}.
Thus an interesting question is whether the power-law of $\rho(\lambda)$ at the critical temperature is consistent with what is expected
from the scaling behavior for the $O(2)$ or $O(4)$ universality class.

In this study, we calculate low-lying eigenvalues of the Dirac operator at several values of the temperature close to the pseudocritical temperature
by using (2+1)-flavor dynamical quarks with the highly improved staggered quark (HISQ) action.
In the following sections, we show the temperature and light quark mass dependence of $\rho(\lambda)$ and its critical behavior.

\section{Simulation setup}

\begin{table}[tbp]
\begin{center}
\begin{tabular}{cccccc}
\hline \hline
$\beta$ & $a$ [fm]& $ T$ [MeV] & $m_s a$ & \multicolumn{2}{c}{\# confs.}   \\ \cline{5-6}
        &           &          &         & $m_l/m_s=1/20$ & $m_l/m_s=1/40$ \\ \hline
6.195   & 0.1847    & 133.5    & 0.0880  & 300            & 142            \\
6.245   & 0.1759    & 140.2    & 0.0830  & 300            & 300            \\
6.260   & 0.1733    & 142.3    & 0.0810  & 300            & 300            \\
6.285   & 0.1691    & 145.8    & 0.0790  & 300            & 300            \\
6.315   & 0.1642    & 150.2    & 0.0760  & 300            & 300            \\
6.341   & 0.1601    & 154.0    & 0.0740  & 300            & 270            \\
6.354   & 0.1581    & 156.0    & 0.0728  & 300            & 300            \\
6.390   & 0.1527    & 161.6    & 0.0694  & 300            & 300            \\
6.423   & 0.1478    & 166.8    & 0.0670  & 300            & 340            \\
6.445   & 0.1447    & 170.5    & 0.0652  & 841            & 246            \\
6.460   & 0.1426    & 173.0    & 0.0640  & 101            &  -              \\
6.488   & 0.1388    & 177.7    & 0.0620  & 183            &  -              \\
6.550   & 0.1307    & 188.7    & 0.0582  & 202            &  -              \\
6.664   & 0.1171    & 210.6    & 0.0514  & 594            &  -              \\
6.800   & 0.1029    & 239.7    & 0.0448  & 599            &  -              \\
6.950   & 0.0894    & 275.9    & 0.0386  & 596            &  -              \\
7.150   & 0.0744    & 331.6    & 0.0320  & 597            &  -              \\
\hline \hline
\end{tabular}
\end{center}
\caption{The parameters of the numerical simulations, i.e. the gauge coupling $\beta$, the lattice spacing $a$, temperature $T$ and the strange quark mass $m_s$ are summarized.
	The number of configurations for measurements is also listed.\label{parameters}}
\end{table}

Our simulations were performed on 32$^3 \times$8 lattices with the tree level improved gauge action and the HISQ action which reduces
the effects of taste symmetry violations and accordingly the cutoff dependence better than the other staggered fermion formulations in use \cite{HISQ}.
Part of our gauge configurations were generated by the HotQCD collaboration \cite{HotQCD}.
The lattice spacing was determined by measuring the static quark anti-quark potential.
The strange quark mass $m_s$ was set to its physical value and the light quark mass $m_l$ was fixed to $m_s/20$ and $m_s/40$ corresponding
to lightest (Goldstone) pions of about 160 and 115 MeV, respectively.
The pseudocritical temperatures for $m_l/m_s=$ 1/20 and 1/40 has been estimated at 162.9(1.8) and 157(3) MeV, respectively, from the peak in the chiral susceptibility.
The details of the determination of the lattice spacing, the strange quark mass and the pseudocritical temperature has been discussed in Ref.~\cite{HotQCD}.
Each 10th trajectory was chosen for measurements after skipping at least 500 trajectories for thermalization.
Statistical errors were estimated by the jackknife method.
Our simulation parameters and statistics are summarized in Table \ref{parameters}.

The staggered Dirac operator $\slash{D}$ is anti-hermitian and it has purely imaginary eigenvalues $i\lambda$. Because of the remnant chiral symmetry,
the eigenvalues always appear in complex conjugate pairs.
We calculated the lowest 100 positive eigenvalues $\lambda_k$ ($k=1,2,\cdots,100$) of $i\slash{D}$, and then
evaluated the eigenvalue density defined by
\begin{equation}
\rho(\lambda)\equiv \frac{1}{V}\langle \sum_k \delta(\lambda-\lambda_k) \rangle \ ,
\end{equation}
where $V$ is the four-volume.
Here $\rho(\lambda)$ is normalized so that $\int d \lambda \rho(\lambda)=$ \# eigenvalues$/V$.
Numerically, $\rho(\lambda)$ was computed by binning of eigenvalues in small intervals for each configuration.
In this study, we chose 0.0005 as the size of a bin for all of the temperatures and light quark masses.

\section{Temperature and light quark mass dependence of $\rho(\lambda)$}
\begin{figure}[tbp]
 \begin{center}
  \includegraphics[width=85mm, angle=-90]{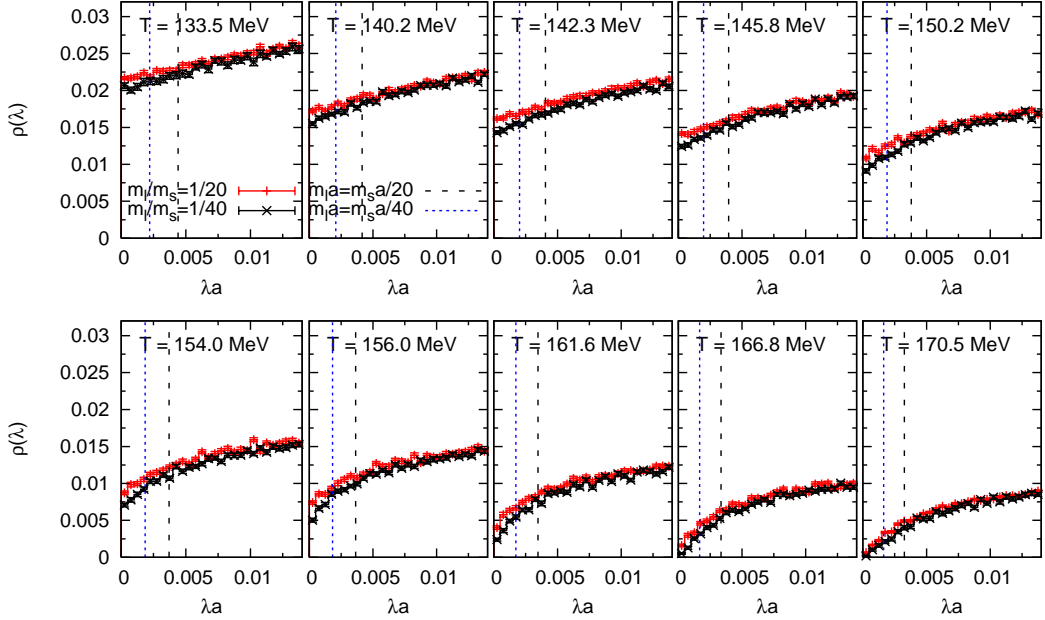}
  \caption{Temperature and light quark mass dependence of the eigenvalue density. Plus and cross symbols indicate $\rho(\lambda)$ for
  $m_l/m_s=1/20$ and $1/40$, respectively. The value of the $m_l$'s are shown by vertical dashed and dotted lines.\label{density_mdep}}
 \end{center}
\end{figure}

Figure \ref{density_mdep} shows the temperature and light quark mass dependence of $\rho(\lambda)$. We found a small quark mass dependence for eigenvalues
$\lambda$ small compared to $m_l$, indicated by vertical dashed and dotted lines in the plots. The bulk behavior of $\rho(\lambda)$, on the other hand, is not sensitive to the quark mass.
Moreover, $\rho(0)$ becomes smaller as the temperature increases and/or the light quark mass decreases, and eventually goes to zero, consistent
with what is expected from the restoration of chiral symmetry, {\it i.e.,} the vanishing of the chiral order parameter above $T_c$.

\begin{figure}[tbp]
 \begin{center}
  \includegraphics[width=52mm, angle=-90]{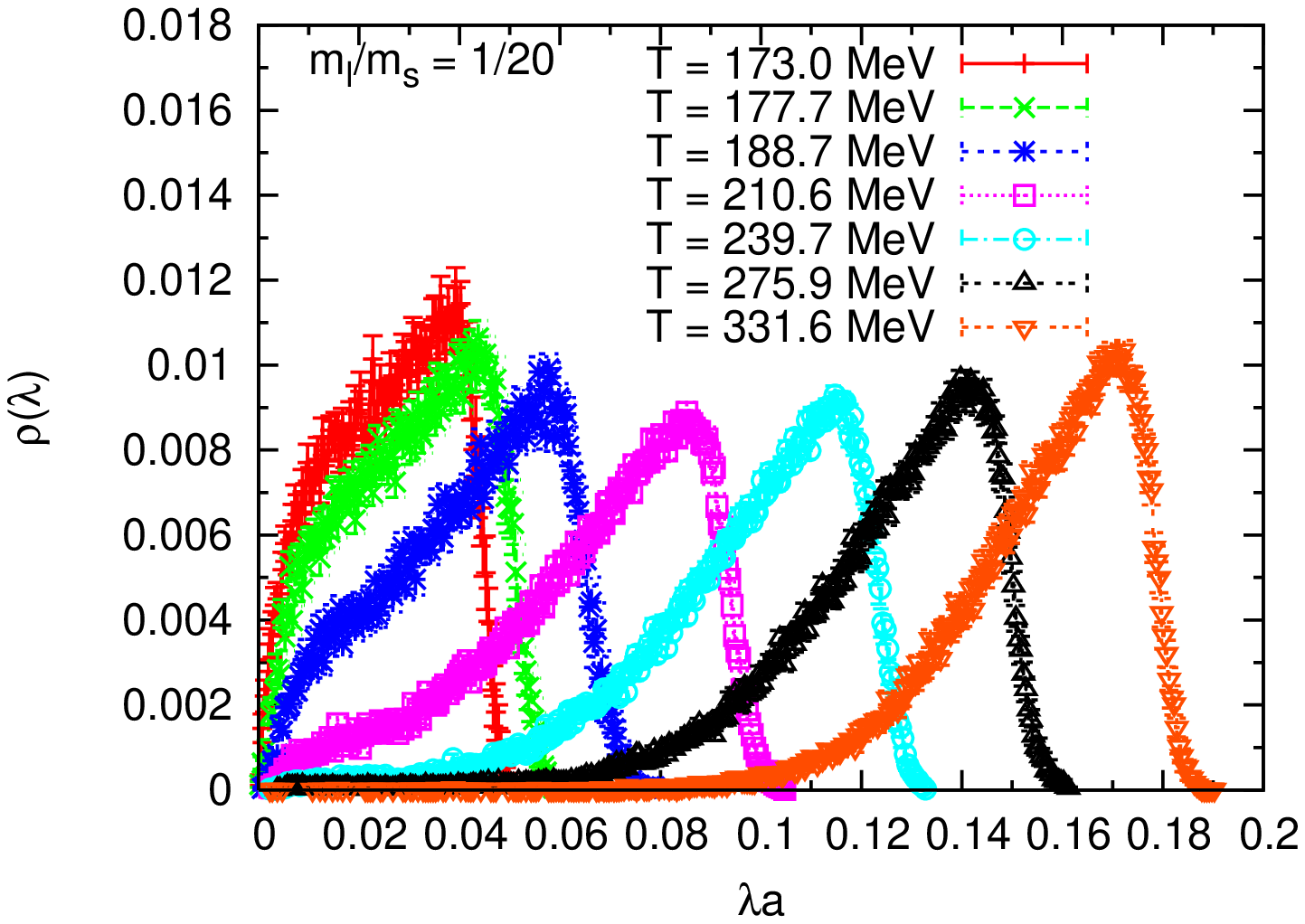}
  \includegraphics[width=52mm, angle=-90]{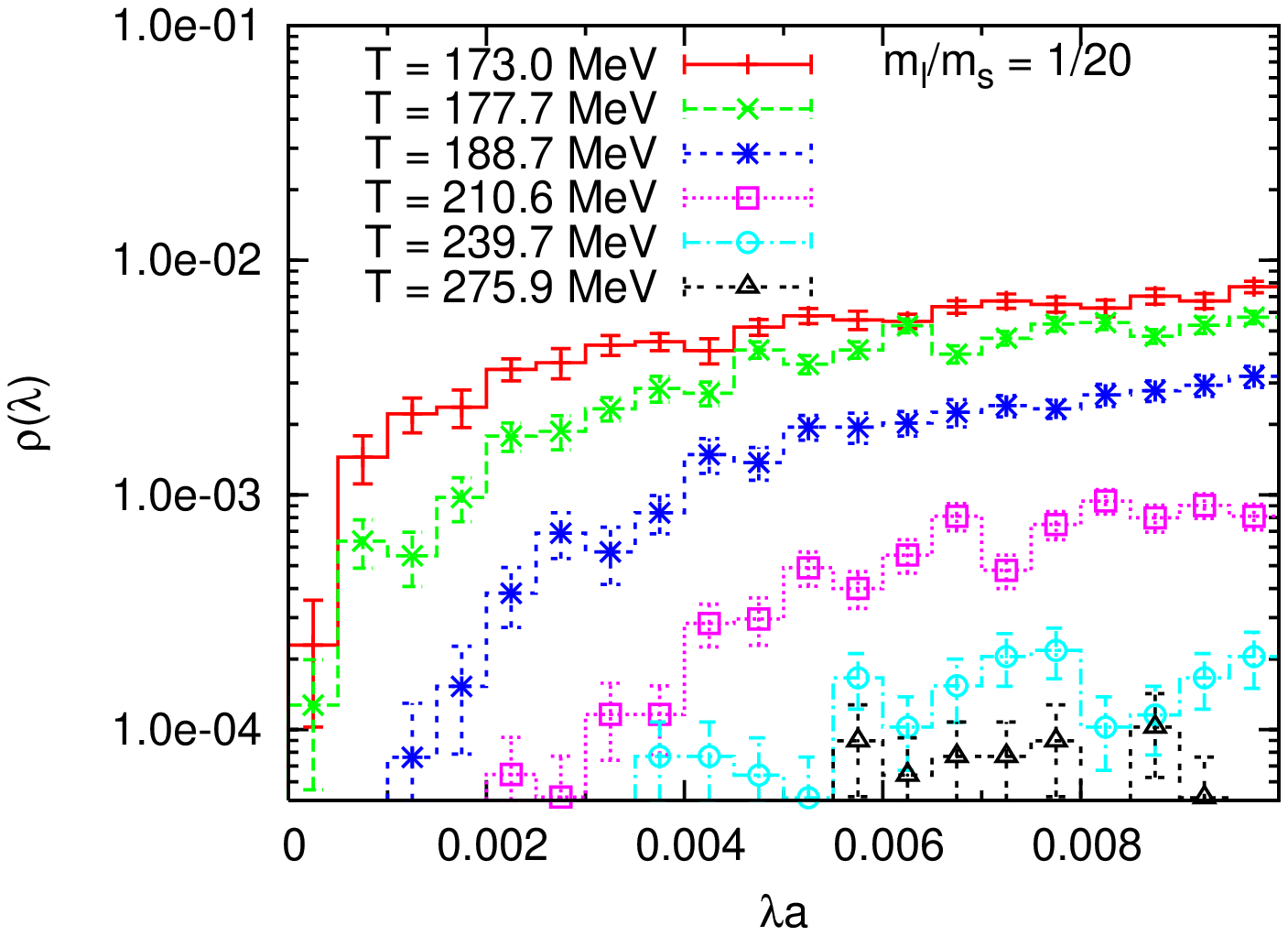}
  \caption{The eigenvalue density above $T_c$. Plus, cross, asterisk, square, circle, triangle, and downward triangle symbols indicate
  $\rho(\lambda)$ at $T=$173.0, 177.7, 188.7, 210.6, 239.7, 275.9, and 331.6 MeV, respectively. The right figure shows a logarithmic plot of
  the same data for the small $\lambda$ region. All of the data points for $T=331.6$ MeV are outside of the region shown.\label{density_highT}}
 \end{center}
\end{figure}

The eigenvalue distribution $\rho(\lambda)$ at temperatures clearly above $T_c$ is shown in Fig.~\ref{density_highT}. The distribution of eigenvalues above $T=239.7$ MeV is qualitatively
different from that below that temperature. Namely, $\rho(\lambda)$ has only a tail that approaches the origin. As mentioned in Sec.~\ref{intro}, the important point here is
whether $\rho(\lambda)$ has a gap around $\lambda=0$. From the r.h.s of Fig.~\ref{density_highT}, a lack of eigenvalues around the origin can be seen
above $T=188.7$ MeV. However, a more careful analysis, including an investigation of the (spatial) volume dependence, is needed to distinguish a real gap from lattice artifacts and
finite volume effects.

\section{Critical behavior of $\rho(\lambda)$}

In the vicinity of a critical point, the behavior of an order parameter $M$ is controlled by a universal scaling function $f_G(z)$ as
\begin{equation}
M=h^{1/\delta}f_G(z) \ ,
\end{equation}
with $z\equiv t/h^{1/{\beta\delta}}$, where $h$ and $t$ are scaling variables corresponding to a symmetry breaking field and temperature, respectively,
and $\beta$ and $\delta$ are critical exponents. In QCD $\langle \bar{\psi}\psi \rangle$ and the (light) quark mass $m$ are regarded as $M$ and $h$, respectively.
Thus one has the relation
\begin{equation}\label{psi_bar_psi}
\langle \bar{\psi}\psi \rangle \sim m^{1/\delta}f_G(z) \ .
\end{equation}

On the other hand, in the infinite volume limit,
$\langle \bar{\psi}\psi \rangle$ can be obtained from the eigenvalues of the Dirac operators as
\begin{equation}\label{psi_bar_psi2}
\langle \bar{\psi}\psi \rangle = -\int^{\infty}_0 d \lambda \frac{2m \rho(\lambda)}{\lambda^2+m^2} \ .
\end{equation}
Assuming $\rho(\lambda) \sim A \lambda^{\alpha}$, Eq.~(\ref{psi_bar_psi2}) can
be rewritten in the limit $m\rightarrow 0$ as
\begin{equation}\label{psi_bar_psi3}
\langle \bar{\psi}\psi \rangle = -m^{\alpha} \int^{\infty}_0 d \bar{\lambda} \frac{2 A \bar{\lambda}^\alpha}{\bar{\lambda}^2+1} \ ,
\end{equation}
with $\bar{\lambda}\equiv \lambda/m$. Thus, by comparing (\ref{psi_bar_psi}) to (\ref{psi_bar_psi3}), $\alpha=1/\delta$ would be expected at $T_c$
in the chiral limit and $\alpha$ should have a value close to $1/\delta$ for a small enough quark mass and near $T_c$.

To test this expectation, we fit the eigenvalue density around $T_c$ to the Ansatz $\rho(\lambda)=A \lambda^{\alpha}$.
Here we set the fit range as $[0,\lambda_{\mathrm{max}}]$.
Since the part of $\rho(\lambda)$ with large $\lambda$ is suppressed due to us having calculated only a fixed number of low-lying eigenvalues per configuration,
the largest $\lambda$ in the region without such a suppression effect is chosen as $\lambda_{\mathrm{max}}$.

\begin{figure}[tbp]
 \begin{center}
  \includegraphics[width=65mm, angle=0]{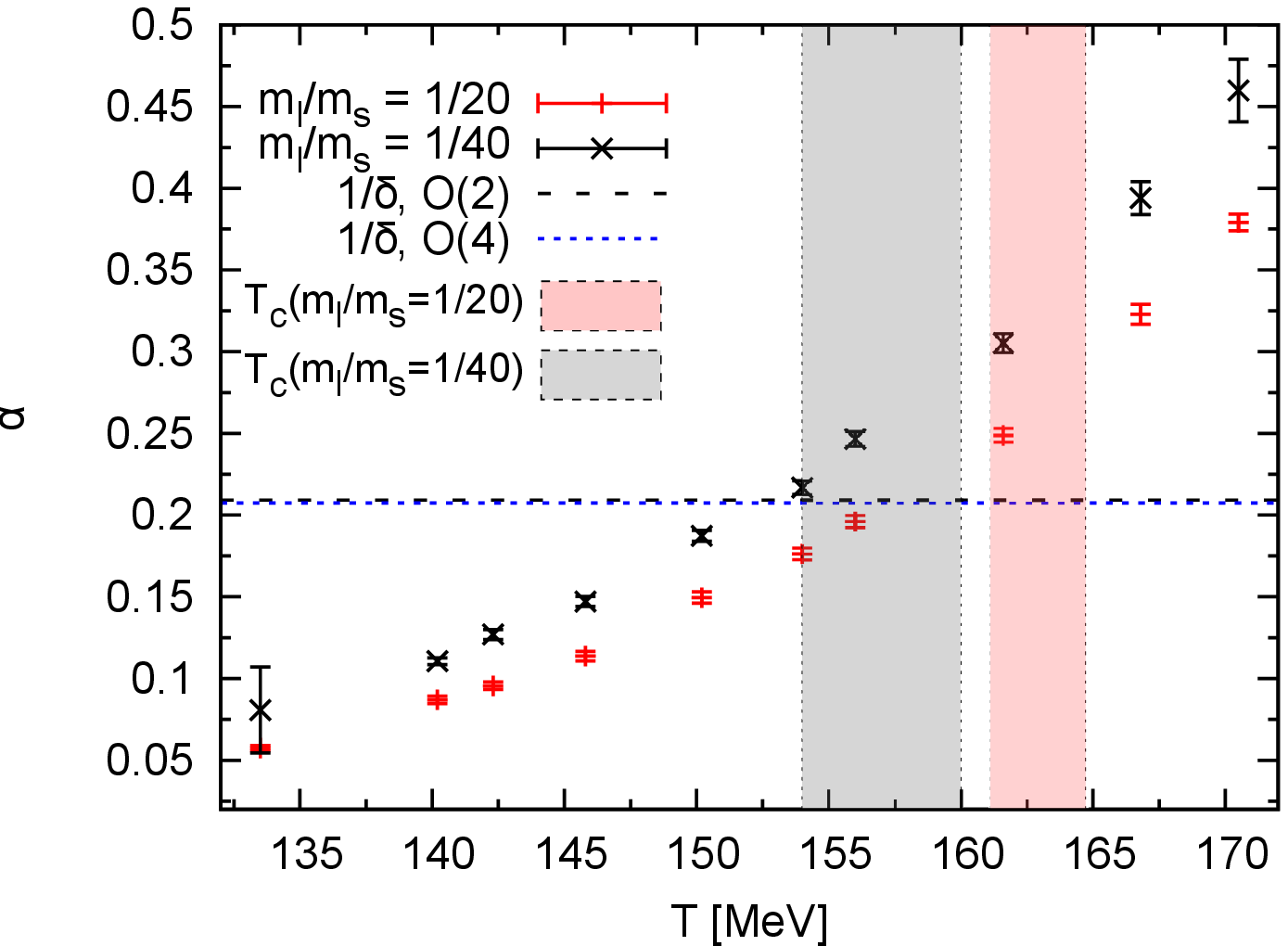} \hspace{1em}
  \includegraphics[width=65mm, angle=0]{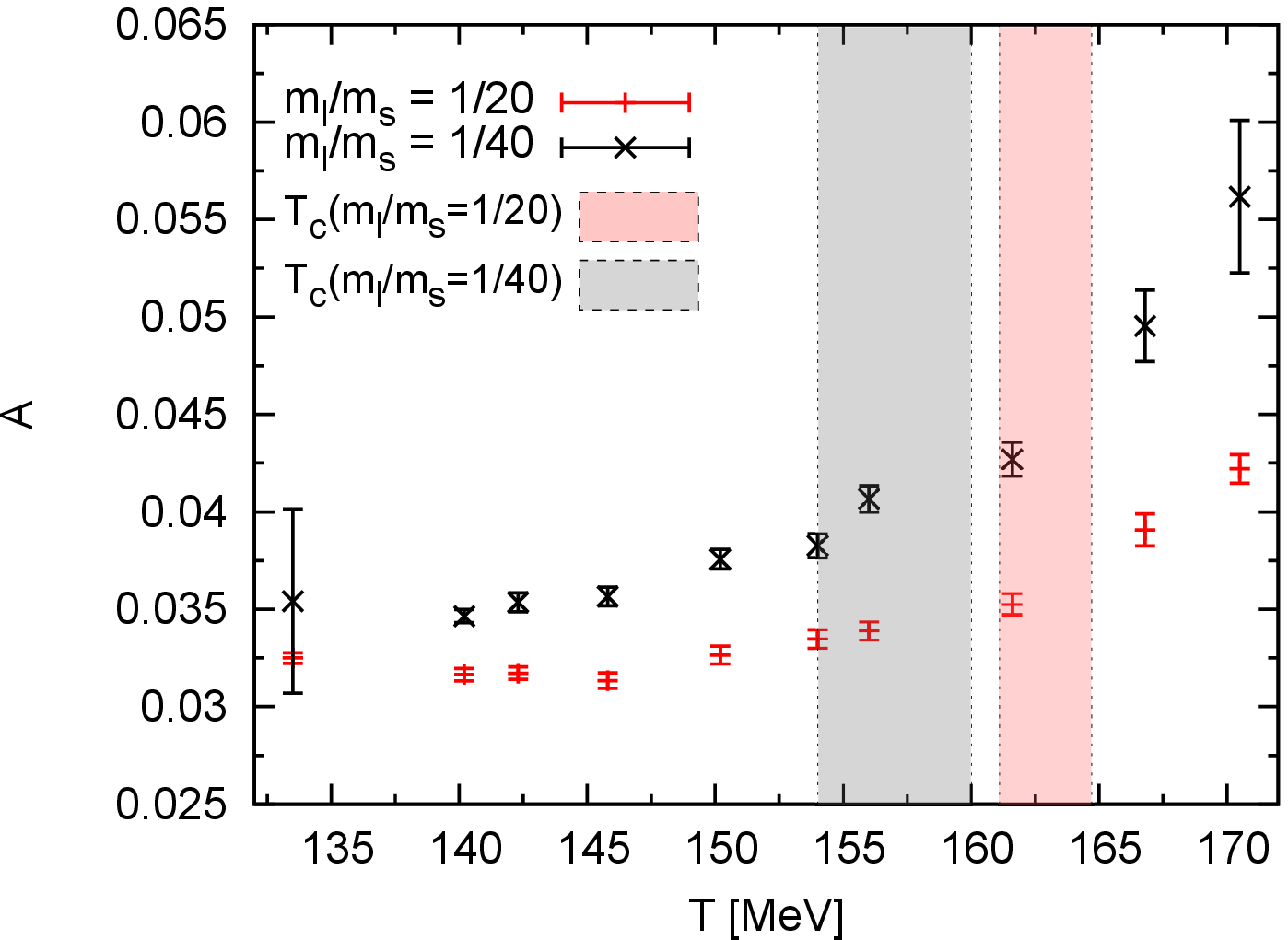}
  \caption{Temperature dependence of the fit parameters $\alpha$ (left) and $A$ (right). Plus and cross symbols correspond to $m_l/m_s=1/20$ and $1/40$ results,
  respectively.  In the left figure, the inverse of the critical exponent $\delta$ for the $O(2)$ \cite{Engels:2001bq} and $O(4)$ \cite{Engels:2003nq} universality classes is indicated by the horizontal dashed and
  dotted lines, respectively. Note that those two horizontal lines are too close to each other to be distinguished.
  The pseudocritical temperature for each light quark mass is shown by a colored band.\label{fit_result}}
 \end{center}
\end{figure}

Figure \ref{fit_result} shows the temperature dependence of the fit parameters $\alpha$ and $A$.
$\alpha$ increases monotonically as the temperature increases and it has a value close to $1/\delta$ for either the $O(2)$ or $O(4)$ universality class\footnote{$1/\delta$ for the $O(2)$
and $O(4)$ universality classes are too similar to be distinguishable within
our numerical accuracy.}
at a temperature not more than 10 MeV below the pseudocritical temperature for both $m_l/m_s=$ 1/20 and 1/40.
Since we expect that $\alpha=1/\delta$ at $T_c$ only in the chiral limit, the
fact that this occurs at a somewhat smaller temperature should be due to the finite $m_l$ used. The deviation becomes smaller as $m_l$ decreases.
We also note that for finite $m_l$, namely in the crossover region, the (pseudo-)critical temperature depends on the quantity which is used to determine it.
$A$ shows a monotonically increasing behavior, too, but it looks insensitive to temperature below $T_c$.

\begin{figure}[tbp]
 \begin{center}
  \includegraphics[width=65mm, angle=0]{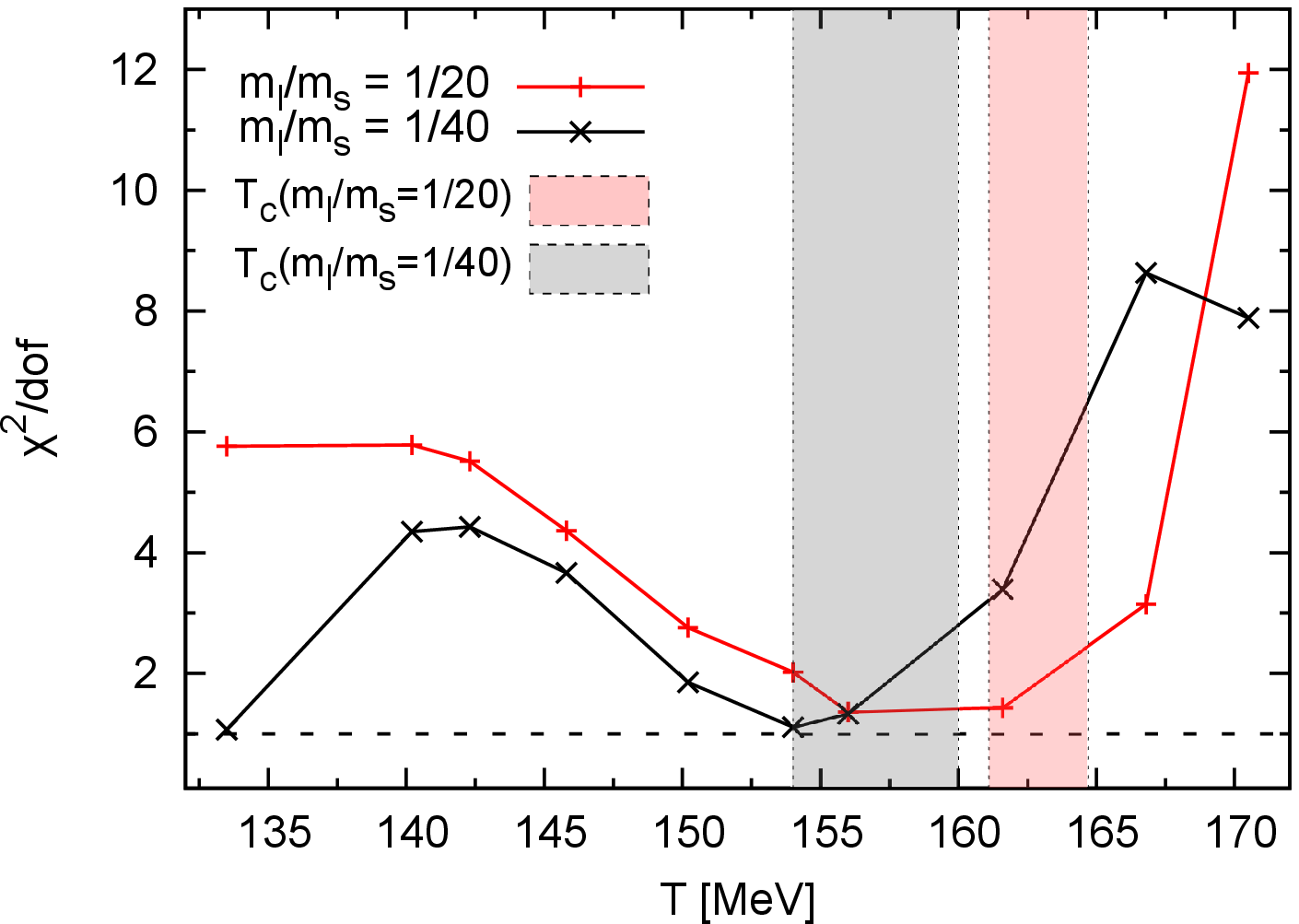} \hspace{1em}
  \includegraphics[width=65mm, angle=0]{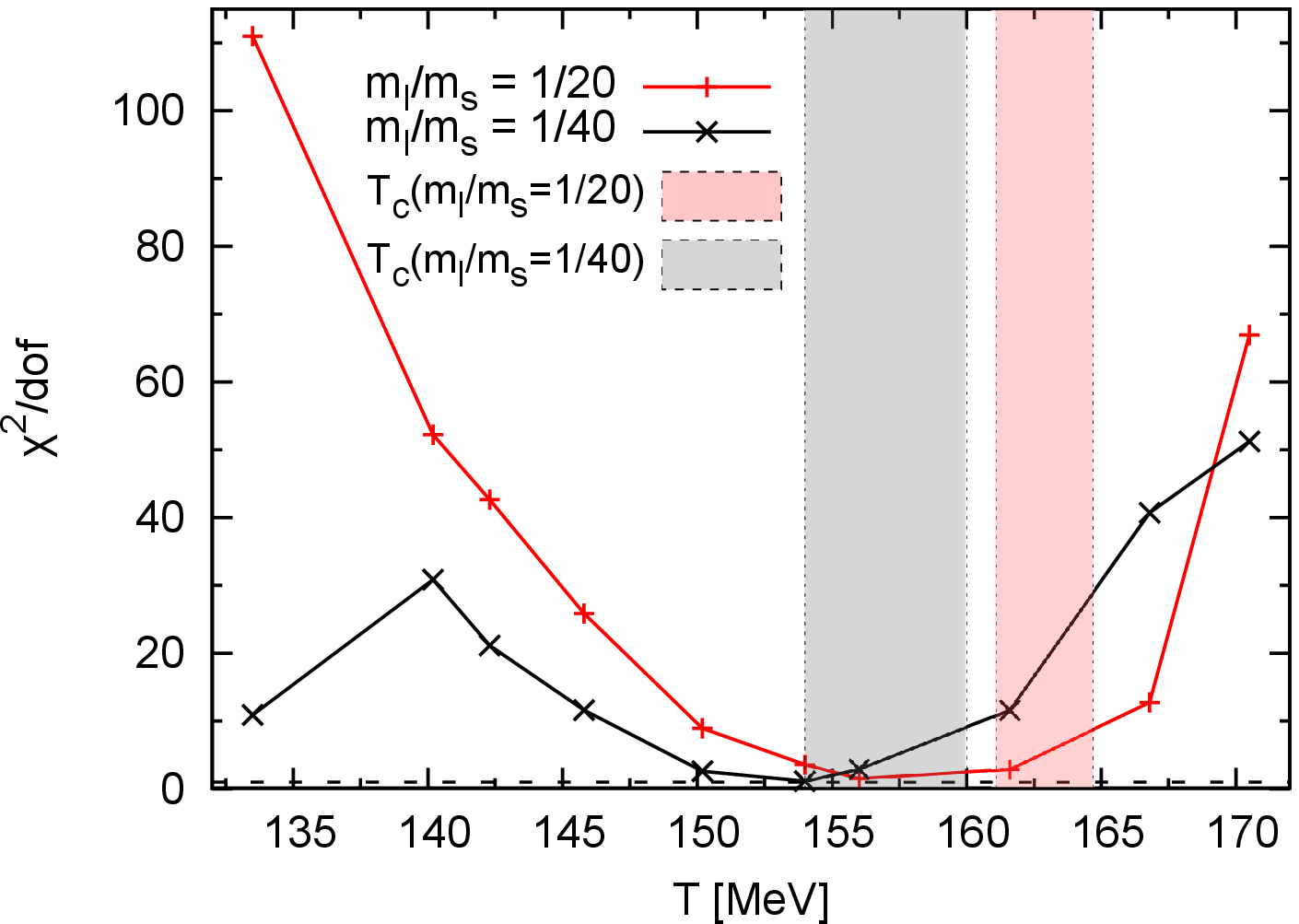}
  \caption{$\chi^2/$dof for our fit analysis with the Ansatz $\rho(\lambda)=A\lambda^\alpha$ at each temperature (left).
  $\chi^2/$dof in the case that $\alpha$ is fixed to $1/\delta$ for the $O(2)$ universality class is shown in the right figure --- note the different y-axis scale.
  The pseudocritical temperature for each light quark mass is shown by a colored band.\label{chisq}}
 \end{center}
\end{figure}

We show the $\chi^2/$dof of our fit analysis in Fig.~\ref{chisq}. We find a minimum of $\chi^2/$dof with a value close to 1
for both of the quark masses near their pseudocritical temperatures. Especially, if we fix $\alpha$ to $1/\delta$ for the $O(2)$
universality class, $\chi^2/$dof is quite large when the temperature is far from the pseudocritical
one\footnote{The plot using the $O(4)$ value of $1/\delta$ would be indistinguishable.}.
This means that our fit analysis and Ansatz work well only around the pseudocritical temperature, which is consistent with what is expected from the scaling behavior of the order parameter
discussed above.

\section{Conclusions}

We studied the behavior of eigenvalue distributions below and above $T_c$ with (2+1)-flavor dynamical quarks using the HISQ action.
We find that the eigenvalue density around $\lambda=0$ decreases as the temperature increases and/or the light quark mass decreases. This is consistent with
what is expected from the behavior of the chiral order parameter. Moreover, we find that the eigenvalue density above $T_c$ has a tail that approaches the origin
and that there is no clear evidence for a gap in the eigenvalue density up to 177.7 MeV, about 1.1$T_c$.

We also investigated the critical behavior around $T_c$ by fitting the eigenvalue density to an Ansatz
$\rho(\lambda)=A\lambda^\alpha$, which is the form expected from the scaling behavior of the order parameter around $T_c$.
We find that the fit analysis works well around the pseudocritical temperature and that the value of $\alpha$ there is compatible with $1/\delta$,
where $\delta$ is the critical exponent for the $O(2)$ or $O(4)$ universality class which governs the behavior of the order parameter ($\langle \bar{\psi}\psi \rangle$) at $T_c$.

\acknowledgments{
HO is supported by the Japan Society for the Promotion of Science for Young Scientists.
FK and SM are supported under Contract No. DE-AC02-98CH10886 with the U.S. Department of Energy.
The numerical simulations have been performed on the BlueGene/L at the New York Center for Computational
Sciences (NYCCS) which is supported by the U.S. Department of Energy and by the State of New York and
the infiniband cluster of USQCD at Jefferson Laboratory.}

\end{document}